\documentstyle[aps,epsfig]{revtex}
\def\b{\bar}
\def\d{\partial}

\def\G{\Gamma}
\def\l{\lambda}

\def\m{\mu}
\def\n{\nu}

\def\t{\tau}

\def\~{\widetilde}
\def\h{\eta}

\def\bY3{\bar Y_{,3}}
\def\Y3{Y_{,3}}
\def\z{\zeta}
\def\Z{{\b\zeta}}
\def\Y{{\bar Y}}

\def\`{\dot}
\def\be{\begin{equation}}
\def\ee{\end{equation}}
\def\bea{\begin{eqnarray}}
\def\eea{\end{eqnarray}}

\def\fn{\footnote}

\def\cF{{\cal F}}

\def\mn{{\mu\nu}}

\begin{document}
\twocolumn

\title{The Kerr theorem and multi-particle Kerr-Schild
solutions}

\author{Alexander Burinskii\\
Gravity Research Group, NSI Russian
Academy of Sciences\\
B. Tulskaya 52, 115191 Moscow, Russia}

\maketitle

\begin{abstract}

We discuss and prove an extended version of the Kerr theorem
which allows one to construct the exact solutions  of the Einstein-Maxwell
field equations from a holomorphic generating function $F$ of twistor
variables. The exact multiparticle Kerr-Schild solutions are
obtained from generating function of the form $F=\prod _i^k F_i, $
where $F_i$  are partial generating functions for the spinning
particles $ i=1...k$. Solutions have an unusual multi-sheeted
structure. Twistorial structures of the i-th and j-th particles do
not feel each other, forming a type of its internal space.
Gravitational and electromagnetic interaction of the particles
occurs via the light-like singular twistor lines. As a result,
each particle turns out to be `dressed' by singular pp-strings
connecting it to other particles. We argue that this solution may
have a relation to quantum theory and hints a geometrical
(twistorial) way to quantum gravity.

\end{abstract}

\bigskip

\section{Introduction}

In the fundamental work by Debney, Kerr and Schild \cite{DKS}, the
Einstein-Maxwell field equations were integrated out for the
Kerr-Schild form of metric

\be g_{\m\n} = \h_{\m\n} + 2h e^3_\m e^3_\n, \label{ksa} \ee

where $\h_{\m\n}$ is metric of an auxiliary Minkowski space-time
$M^4,$ and vector field $e^3_\m (x)$ is null
($e^3_\m e^{3\m} =0$) and tangent to a
 principal null congruence (PNC) which is geodesic and shear-free (GSF)
\cite{DKS}. PNC is determined by a complex function $Y(x)$  via
the one-form
\be
 e^3 = du+ \Y d \z  + Y d \Z - Y \Y d v
\label{cong} \ee
written in the null Cartesian coordinates\fn{Coordinates
$x=x^\m$ are Cartesian coordinates of
Minkowski space $x=\{x,y,z,t \}\in M^4 .$ It is assumed that they
may be analytically extended to a complexified Minkowski space
$CM^4$. The function $Y$ is a projective angular coordinate,
i.e. projection of sphere $S^2$ on a complex plane.}
\bea
2^{1\over2}\z &=& x + i y ,\qquad 2^{1\over2} \Z = x - i y , \nonumber\\
2^{1\over2}u &=& z + t ,\qquad 2^{1\over2}v = z - t . \label{ncc}
\eea

One of the most important solution of this class is the
Kerr-Newman solution for the rotating and charged black hole. It
has been mentioned long ago that the Kerr-Newman solution displays
some relationships to the quantum world. It has the anomalous
gyromagnetic ratio $g=2$, as that of the Dirac electron
\cite{Car}, stringy structures \cite{IvBur1,BurStr,BurOri,BurTwi} and
other features allowing one to construct a semiclassical model of
the extended electron \cite{BurTwi,Isr,Bur0,IvBur,Lop,New1} which
has the Compton size and possesses the wave properties
\cite{BurTwi,Bur0,BurPra}. So, we will be speaking on {\it the
Kerr spinning particle}, assuming  a particle-like source
of the Einstein -Maxwell field equations generating the external
Kerr-Newman field and having a definite
position of centrum of mass, and the definite momentum and
orientation of angular momentum.

 Principal null congruence (PNC) of the Kerr-Newman solution
represents a vortex of the light-like rays (see Fig.1) which are
{\it twistors} indeed. So, the Kerr geometry is supplied by a {\it
twistorial structure}  which is described in twistor terms by the
Kerr theorem. In addition to the very important meaning for
twistor theory \cite{Pen,PenRin,KraSte}, the Kerr theorem
 represents in the Kerr-Schild approach \cite{DKS} {\it a very
useful technical instrument} allowing one to obtain the
Kerr-Newman solution and its generalizations.

 In accordance with the  Kerr theorem, the
general geodesic and shear-free congruence on $M^4$ is generated
by the simple algebraic equation \be F = 0 , \label{KT}\ee where
$F(Y,\l_1,\l_2)$ is any holomorphic function of the
projective twistor coordinates \be Y,\quad \l_1 = \z - Y v, \quad
\l_2 =u + Y \Z .\label{Tw}  \ee Since the twistor coordinates
$\l_1,\l_2$ are itself the functions of $Y$, one can consider $F$
as a function of $Y$ and $x\in M^4 $, so the solution of
(\ref{KT}) is a function $Y(x)$ which allows one to restore PNC by
using the relation (\ref{cong}).

We shall call the function $F$ as {\it generating function of the
Kerr theorem.}

It should be noted that the Kerr theorem has never been published
by R. Kerr as a theorem. First, it has been published without a
proof in the Penrose work ``Twistor Algebra'' \cite{Pen}.
However, in a restricted form\fn{For a special type of
generating function $F$.} it has practically been used in
\cite{DKS} by derivation of the Kerr-Newman solution.
The text of paper \cite{DKS} contains some technical details
which allows one to reconstruct the proof of the Kerr theorem
in a general form which is  valid for the Kerr-Schild
class of metrics \cite{BurNst,BurKer}.
 We reproduce this proof in the Appendix B.

The basic results of the fundamental work
\cite{DKS} were obtained for the quadratic in $Y$
generating function $F$, which corresponds to the
Kerr PNC. In particular, for the
Kerr-Newman solution the equation $F(Y)=0$ has two roots $Y^\pm
(x)$ \cite{IvBur1,KerWil,BurNst}, and the space-time is double
sheeted, which is one of the mysteries of the Kerr geometry, since
the $(+)$ and $(-)$ sheets are imbedded in the same Minkowski
background having dissimilar gravitation (and electromagnetic)
fields, and the fields living on the $(+)$-sheet do not feel the
existence of different fields on the $(-)$-sheet.

It has been mentioned in \cite{BurMag}, that for quadratic in $Y$
functions $F$ the Kerr theorem determines not only congruence,
but also allows one  to determine the metric and electromagnetic
field (up to an arbitrary function $\psi(Y)$).

In this paper we give the proof of an
{\it extended version of the Kerr theorem} which
 allows one to determine
the corresponding geodesic and shear-free PNC for a very broad
class of  holomorphic generating functions $F$, and also to
reconstruct the metric and electromagnetic field, i.e.
to describe fully corresponding class of the exact solutions of
the Einstein-Maxwell field equations.

In particular,  we consider polynomial generating
functions $F$ of higher degrees in $Y$ which lead to the
multiparticle Kerr-Schild solutions.
These solutions have a new peculiarity: the space-time and
corresponding twistorial structures turn out to be multi-sheeted.

The wonderful twosheetedness of the usual Kerr space-time is
generalized in these solutions to multi-sheeted space-times
which are determined by  multi-sheeted Riemann holomorphic
surfaces and induce the corresponding
multi-sheeted twistorial structures.

Twistorial structures of the i-th and
j-th particles do not feel each other,
forming a type of its internal space. However, the corresponding
exact solutions of the Einstein-Maxwell field
equations show that particles interact via the common singular twistor
lines -- the light-like pp-strings.
The appearance of singular strings is typical for the known multiparticle
solutions \cite{KinWal}.  However, contrary to the usual cases,
these strings  are light-like and do not have conical
singularities.

We find out that the mystery of the known two-sheetedness of the Kerr
geometry is generalized to some more mystical multi-sheetedness of the
multiparticle solutions.

We find out that besides the usual Kerr-Newman solution for an isolated
spinning particle, there is a series of the exact solutions, in which the
selected Kerr-Newman particle is surrounded by external particles and
interacts with them by singular pp-strings. It is reminiscent of
the known from quantum theory difference between the ``naked''
one-particle electron of the Dirac theory and a multi-particle structure of a
``dressed'' electron which is surrounded by virtual photons in accordance
with QED.
The multiparticle space-time turns out to be penetrated by a
multi-sheeted web of twistors.
Since they are not observable at the classical level, we
conjecture that these multi-sheeted twistor
congruences are related to a twistorial structure of vacuum and
may be a substitute for the usual spin-foam of quantum gravity,
in the spirit of the spin net-work  proposals by Penrose.
Taking also into account the other known relations of the Kerr
geometry to quantum theory, one can conjecture that multisheeted
Kerr geometry gives a hint for a new way to quantum gravity.

Basically, we use the notations of the fundamental paper by Debney, Kerr
and Schild \cite{DKS}, however, the
the Kerr-Schild form of metric (\ref{ksa})
is very convenient for derivation, but for
applications an another equivalent form is more convenient

\be g_{\m\n} = \h_{\m\n} + 2H k_\m k_\n . \label{ksnorm} \ee
 It is obtained
by a simple replacement $h=HP^{-2}, \quad e^3_\m = P k_\m,$
where $P=P(Y,\bar Y)$ is a function which normalizes the vector
field $e^3 _\m$, removing singularity by $Y\to \infty$.
Function $P$ is also determined by the Kerr Theorem.

For the reader
convenience we give in the Appendix A a  brief description
of the basic relations of the Kerr-Schild formalism. In the
Appendix B we reproduce the derivation of the Kerr theorem for
the Kerr-Schild backgrounds, following to the papers
\cite{BurNst,BurKer}. In the Appendix C we describe briefly
the results of \cite{DKS},  the basic
Kerr-Schild equations for the general geodesic and shearfree
congruences.
Appendix D is written for the mathematically oriented readers and
contains a description of the double
twistor bundle forming the geodesic and shearfree congruences
on the Minkowski and the Kerr-Schild space-times.

\section{The Kerr Theorem and one-particle Kerr-Schild solutions}

The Kerr-Schild form of metric (\ref{ksa}) has the remarkable
properties allowing one to apply {\it rigorously} the Kerr Theorem
to the curved spaces.  It is related to the fact that the PNC field
$e^3_\m$, being null and GSF with respect to the Kerr-Schild metric $g$,
$e^{3\m} e^3 _\m |_g= 0,$ will also be null and GSF with respect to
the auxiliary Minkowski metric,
and this relation remains valid by an analytic
extension to the complex region.
\fn{Indeed $e^{3\m} e^3 _\m |_g =e^{3\m} e^3 _\m |_\eta +
 2h (e^{3\m} e^3 _\m |_g)^2 ,$
 which yields $e^{3\m} e^3 _\m |_\eta=0 .$ And vice versa,
 it follows from $e^{3\m} e^3 _\m |_\eta=0$  two solutions for
 $e^{3\m} e^3 _\m |_g,$ one of which corresponds to definition of
 the Kerr-Schild metric, i.e. $e^{3\m} e^3 _\m |_g=0$.}
 In the Appendix A
we show that the geodesic and shear-free conditions on PNC
coincide in the Minkowski space and in the Kerr-Schild background.
 Therefore,
obtaining a geodesic and shear-free PNC in Minkowski space in
accordance with the Kerr theorem, and using the corresponding null
vector field $e^3_\m(x)$ in the Kerr-Schild form of metric, one
obtains a curved Kerr-Schild space-time where PNC will also be
null, geodesic and shearfree.

\begin{figure}[ht]
\centerline{\epsfig{figure=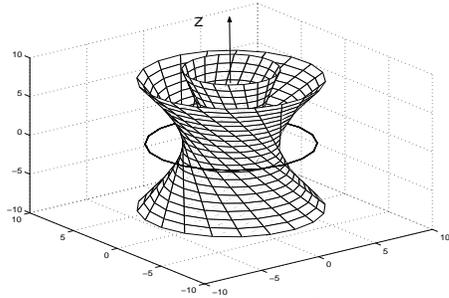,height=4cm,width=6cm}}
\caption{The Kerr singular ring and 3-D section of the Kerr
principal null congruence. Singular ring is a branch line of
space, and PNC propagates from ``negative'' sheet of the Kerr
space to ``positive '' one, covering the space-time twice. }
\end{figure}

It was shown in \cite{BurNst,BurMag} that the quadratic in $Y$
generating function of the Kerr theorem can be expressed via a
set of parameters $q$ which determine the position, motion and orientation
of the Kerr spinning particle.

For some selected particle $i$, function $F_i(Y)$, may be represented in
general form
\be F_i(Y|q_i)=A(x|q_i)Y^2 +B(x|q_i)Y +C(x|q_i)
\label{FiABC}. \ee

The equation $F_i(Y|q_i)=0$ can be resolved explicitly, leading to two
roots $Y(x)=Y^\pm (x|q_i)$ which correspond to two sheets of the
Kerr space-time. The root $Y^+(x)$ determines via (\ref{cong}) the
out-going congruence  on the $(+)$-sheet, while the root $Y^-(x)$
gives the in-going congruence on the $(-)$-sheet.
By using these root solutions, one can represent function
$F_i(Y)$ in the form
\be
F_i(Y) = A_i (x)(Y - Y_i^+(x)) (Y - Y_i^-(x))
\label{iblock}. \ee

The relation (\ref{cong})  determines the vector field
$e^{3(i)}_\m$ of the Kerr-Schild ansatz  (\ref{ksa}), and
metric acquires the form
\be g^{(i)}_\mn =\eta _\mn + 2h^{(i)} e^{3(i)}_\m e^{3(i)}_\n
\label{gi} .\ee

Based on this ansatz, after rather long calculations and
integration of the Einstein-Maxwell field equations {\it performed
in the work \cite{DKS}} under the
conditions that PNC is geodesic and shearfree (which means
$Y,_2=0$  and $Y,_4 = 0, $ see Appendix A), one can
represent the function $h^{(i)}$ in the form
\be
h^{(i)}=\frac 12 M^{(i)}(Z^{(i)}+\bar Z ^{(i)}) -
\frac 12 A^{(i)}\bar A^{(i)} Z^{(i)} \bar Z^{(i)}   ,\label{hi}
\ee

where
\be M^{(i)}=m^{(i)}(P^{(i)})^{-3} \label{Mi}
\ee
and
\be A^{(i)}=\psi^{(i)}(Y) (P^{(i)})^{-2}.\label{Ai}
\ee
Here $m^{(i)}$ is mass and $\psi^{(i)}(Y)$ is arbitrary
holomorphic function.

Electromagnetic field is determined by two complex self-dual
components of the Kerr-Schild tetrad form
$\cF=\cF_{ab} e^a\wedge e^b,$

\be
\cF ^{(i)} _{12} =A^{(i)}( Z^{(i)} )^2 \label{F12i}
\ee
and
\be
\cF ^{(i)} _{31} = -(A^{(i)}Z^{(i)}),_1 \label{F31i},
\ee

see Appendix C.

We added here the indices $i$ to underline that these functions
 depend on the parameters $q_i$ of $i$-th particle.

Setting $\psi^{(i)}(Y)=e=const.$ we have the charged Kerr-Newman
solution for i-th particle, vector potential of which may be
represented in the form

\be {\cal A}_\m^{(i)} =\Re e (e Z^{(i)}) e^{3(i)}_\m
 (P^{(i)})^{-2}  \label{Ami}.\ee

{\bf It should be emphasized (!)},
that integration of the field equations
 has been performed in \cite{DKS} in a general form,
{\it before concretization of the form of congruence,}
only under the general conditions that PNC is geodesic and
shear free.

On the other hand, it was shown in \cite{BurNst,BurMag}
(see also Appendix B), that
the unknown so far functions $P^{(i)}$ and $Z^{(i)}$ can also be
determined from the generating function of the Kerr
theorem $F^{(i)}.$ Namely,
\be P^{(i)} = \d_{\l_1} F_i - \Y \d_{\l_2} F_i , \quad
P^{(i)}/Z^{(i)}= - \quad d F_i / d Y  \   \label{PF} .\ee

Therefore, for the quadratic in $Y$ functions (\ref{FiABC}),
we arrive at the first {\bf extended version of the Kerr theorem}

 {\it 1/ For a given quadratic in $Y$ generating function $F_i$,
 solution of the equation $F_i=0$ determines the geodesic and
 shear free PNC in the Minkowski space $M^4$ and in the
 associated Kerr-Schild background (\ref{gi}).

      2/ The given function $F_i$ determines the exact
 stationary solution of the Einstein-Maxwell field equations
 with  metric given by (\ref{gi}), (\ref{hi}), (\ref{Mi}) and
 electromagnetic field given by  (\ref{Ami}), where functions
 $P^{(i)}$   and $Z^{(i)}$   are given by  (\ref{PF}). }

As we have mentioned in introduction, for practical calculations
the Kerr-Schild form (\ref{ksnorm}) is  more useful, where
function $H=H^{(i)} =h^{(i)} (P^{(i)})^2, $ and the normalized null vector
field is
$k^{(i)}_\m = e^{3(i)}_\m /P^{(i)}$.

In this form function $H^{(i)}$ will be

\be H^{(i)} = \frac m2 (\frac 1 {\tilde r _i} + \frac 1 {\tilde
r_i^*}) + \frac {e^2}{2 |\tilde r_i|^2} \label{Hi}, \ee
and the Kerr-Newman  electromagnetic field is determined by the
vector potential
\be  A_\m^{(i)} =\Re e (e/\tilde r _i) k^{(i)}_\m \label{Aisol}
,\ee
where

\be \tilde r_i =P^{(i)}/Z^{(i)}= - \quad d F_i / d Y  \
\label{tilderi} \ee
is the so called  {\it complex radial distance} which is related to
a complex representation of the Kerr geometry
\cite{BurNst,BurKer,BurMag,DirKer}.

For a standard oriented Kerr
solution in the rest, $\tilde r =\sqrt{x^2 +y^2 + (z-ia)^2}
=r+ia\cos\theta ,$ which corresponds to the distance from a
{\it complex point} source positioned at the
complex point $\vec x =(0,0,ia).$ One sees, that the Kerr
singular ring is determined by
$\tilde r =0\Rightarrow r=\cos \theta =0$.
For the Kerr geometry this
representation was initiated by Newman, however it works
rigorously in the Kerr-Schild approach,
where the complex source represents a complex world line $x(\t)$
in the complexified auxiliary Minkowski space-time $CM^4$.
We will use this approach in sec.5.

In accordance with Corollary 4 (Appendix B), position of the
Kerr singular ring is determined by the system of equations
\be
F_i=0, \quad \tilde r_i = - \quad d F_i / d Y =0  \
\label{singi} .\ee

Extended version of the Kerr theorem
allows one to get exact  solution for an arbitrary oriented and
boosted charged spinning particle \cite{BurMag}.

\section{Multi-sheeted twistor space}

Let us now consider the case of a system of $k$ spinning particles having
the arbitrary displacement, orientations and boosts.
One can form the function $F$ as a product of the corresponding blocks
$F_i(Y)$,

\be F(Y)\equiv \ \prod _{i=1}^k F_i (Y) \label{multi}. \ee

The solution of the equation $F=0$ acquires $2k$ roots
$Y_i^\pm(x),$ forming a multisheeted covering space over the
Riemann sphere $S^2=CP^1\ni Y$.

Indeed, $Y=e^{i\phi}\tan\frac \theta 2$ is a complex projective
angular coordinate on the Minkowski space-time and on the
corresponding Kerr-Schild space-time.\fn{Two other coordinates
in the Kerr-Schild space-time may be chosen as $\tilde r =PZ^{-1}$
 and $\rho=x^\m e^3_\m ,$ where $x^\m$ are the four Cartesian
coordinates in $M^4$.}

\begin{figure}[ht]
\centerline{\epsfig{figure=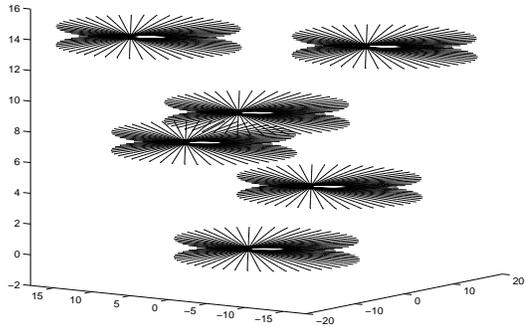,height=5cm,width=7cm}}
\caption{Multi-sheeted twistor space over the auxiliary Minkowski
space-time of the multiparticle Kerr-Schild solution. Each
particle has two-sheeted structure.}
\end{figure}

The twistorial structure on
the i-th $(+)$ or $(-)$ sheet is determined by the equation
$F_i=0$
and does not
depend on the other functions $F_j , \quad j\ne i$. Therefore, the
particle $i$ does not feel the twistorial structures of other
particles.

The equations for singular lines
\be F=0, \ dF/dY =0 \label{singlines}\ee
 acquires the form
\be \prod _{l=1 }^k F_l =0, \qquad   \sum ^k_{i=1} \prod _{l\ne i}^k
F_l d F_i/dY =0 \label{leib} \ee

which splits into k independent relations

\be F_i=0,\quad \prod
_{l\ne i}^k F_l d F_i /dY=0 \label{kind}. \ee

One sees, that the Kerr singular ring on the sheet i  is
determined by the usual relations $F_i=0, \quad d F_i /dY=0,$
and i-th particle does not feel also the singular
rings of the other particles.
The  space-time splits on the independent
twistorial sheets, and  the twistorial structure related
to the i-th particle plays the role of  its ``internal space''.
One should mention that it is a direct generalization of the
well known two-sheetedness of the usual Kerr space-time.

Since the twistorial structures of different particles are
independent, it seems that  the k-particle solutions $\{ Y^\pm_i
(x)\}, \ i=1,2..k$ form a trivial covering space $K$ over the
sphere $S^2$, i.e. $K$ is a trivial sum of k disconnected
two-sheeted subspaces $K=\bigcup ^k_i S^2_i$.

However, there is one more source of singularities on $K$ which
corresponds to the multiple roots: the cases when some of twistor
lines of one particle $i$ coincides with a twistor line of another particle $j$,
forming a common $(ij)$-twistor line. Indeed, for each pair of
particles $i$ and $j$, there are two such common twistor lines:
one of them $(\vec{ij})$ is going from the positive sheet of
particle $i$, $Y_i^+(x)$ to negative sheet of
particle $j$, $Y_j^-(x)$ and corresponds to the solution of
the equation $Y_i^+(x)=Y_j^-(x),$ another one
$(\vec{ji})$ is going from the positive sheet of
particle $j$, $Y_j^+(x)$ to negative sheet of
particle $i$ and corresponds to
the equation $Y_j^+(x)=Y_i^-(x).$ We will consider the corresponding
simple example in sec.5.

The common twistor lines are also
described by the
solutions of the equations (\ref{singlines}) and correspond
to the multiple roots which give a set of ``points'' $A_j,$
where the complex analyticity of the map $Y^\pm_i(x)\to S^2$ is
broken.\fn{The given
 in \cite{DNF} analysis of the equations (\ref{singlines})
shows that for the holomorphic functions $F(Y)$
 the covering space K turns out to be connected and forms
 a  multisheeted Riemann surface
over the sphere  with the removed branch points
$S^2 \setminus \cup_j A_j .$}

The solutions $Y_i(x),$ which determine PNC on the i-th sheet
of the covering space, induce
multisheeted twistor fields over the corresponding
Kerr-Schild manifold ${\cal K}^4 .$

\section{Multiparticle Kerr-Schild solutions.}

As we have seen,  the quadratic in $Y$ functions $F$  generate
exact solutions of the  Einstein-Maxwell field equations.
In the same time, the considered above generating
functions $\prod _{i=1}^k F_i (Y)=0,$ leads to a multisheeted
covering space over $S^2$ and to the induced multisheeted twistor
structures over the Kerr-Schild background which look like
independent ones. Following to the initiate naive assumption
that twistorial sheets are fully independent, one could expect
that the corresponding multisheeted solutions of the
Einstein-Maxwell field equations will be
independent on the different sheets, and the solution
on i-th sheet  will reproduce the result for an isolated
i-th particle.
However, It is obtained that the result is different.

Formally, we have to replace $F_i$ by
\be F=\prod _{i=1}^k F_i
(Y)=\mu _i F_i(Y) ,\label{Fk}\ee
where
\be\mu _i =\prod _{j\ne i}^k F_j (Y)
\label{mui}\ee
 is a normalizing factor which  takes into account the external
 particles.
In accordance with (\ref{PF}) this factor will also
appear  in the new expression for $P/Z$ which we
mark now by capital letter $\tilde R_i$
\be
\tilde R_i = P/Z =- d_Y F= \mu _i P^{(i)}/Z^{(i)} , \label{Ri}
\ee
and in the new function $P_i$ which we will mark by hat
\be
\hat P_i  = \mu_i P_i  . \label{hPi}
\ee
Functions $Z$ and $\bar Z$ will not be changed.

By substitution of the new functions $P_i$ in the
relations (\ref{hi}), (\ref{Mi}) and (\ref {Ai}),
we obtain the new relations

\be M^{(i)}=m^{(i)}(\m_i(Y) P^{(i)})^{-3} ,\label{NMi}
\ee
\be A^{(i)}=\psi^{(i)}(Y) (\m_i(Y) P^{(i)})^{-2}\label{NAi}
\ee
and
\be h_i=\frac {m} {2(\m _i (Y) P_i)^3} (Z^{(i)} + \bar Z^{(i)}) - \frac {|\psi|^2}
{2|\m_i(Y)P_i|^4} Z^{(i)}\bar Z^{(i)} \label{hmudks}.  \ee

For new components of electromagnetic field we obtain
\be
\cF ^{(i)} _{12} =\psi^{(i)}(Y) (\m_i P^{(i)})^{-2}
(Z^{(i)})^2 \label{NF12i}
\ee
and
\be
\cF ^{(i)} _{31} = -(\frac {\psi^{(i)}(Y)} {(\m_i(Y) P^{(i)})^{2}}
Z^{(i)}),_1 \label{NF31i}.
\ee

In the terms of $\tilde r_i$ and $H_i$ the Kerr-Newman metric
takes the form:

\be H_i = \frac {m_0}2 (\frac 1{\mu_i\tilde r_i} + \frac 1 {\mu_i^*\tilde
 r_i^*}) + \frac {e^2}{2 |\mu_i \tilde r_i|^2} , \label{hKSi} \ee

where we have set $\psi^{(i)}(Y)= e, $ for the Kerr-Newman solution.

The simple expression for vector potential  (\ref{Ami}) is not valid
more.\fn{Besides the related with $\m_i(Y)$ singular string factor, it
acquires an extra vortex term.}

One sees, that in general case metric turns out to be complex
for the complex mass factor $ m (\m _i (Y))^{-3}$, and one has to
try to reduce it to the real one.

This problem of reality $M$ was also considered in \cite{DKS}.
Function $M$ satisfies the equation
\be (\ln M + 3\ln P),_\Y =0 \ee
which has the general solution
\be M=m(Y)/P^3(Y,\bar Y) ,\label{mY} \ee
where $m(Y)$ is an arbitrary holomorphic function.
The simplest real solution is given by
the real constant $m$ and the real function $P(Y,\bar Y)$.
As it was shown in \cite{DKS}, it resulted in the one-particle
solutions.

In our case functions $P_i$ have also to be real, since they
relate the real one-forms $e^3$ and $k$,
\be e^{3(i)} =P_i k^{(i)}_\m dx. \label{ePk}
\ee
Functions $\m _i(Y)$ are the holomorphic
functions given by (\ref{mui}), and functions $m_i=m_i(Y)$ are
arbitrary holomorphic functions which may be taken in the form
\be
m_i(Y)=m_0 (\m _i(Y) )^3
\ee
to  provide reality of the mass terms $M_i=m_i(Y)/{\hat P_i}^3 $
on the each i-th sheet of the solution.

Therefore, we have achieved the reality of the multisheeted
Kerr-Schild solutions, and
{\it the extended version of the Kerr theorem
is now applicable for the general multiplicative form of the
functions $F$, given by (\ref{multi}).}

One can specify the form of functions $\mu_i$
by using the known structure of  blocks $F_i$
\be \mu_i (Y_i)=
\prod _{j\ne i} A_j (x)(Y_i - Y_j^+) (Y_i - Y_j^-)
\label{muYi}. \ee
If the roots $Y^\pm_i$ and $Y_j^\pm$ coincide for some values of
$Y^\pm_i$,
it selects a common twistor for the sheets $i$ and $j$. Assuming
that we are on the i-th $(+)$-sheet, where congruence is out-going,
this twistor line will also  belong to the in-going $(-)$-sheet of
the particle $j$ . The metric and electromagnetic  field will be
singular along this twistor line, because of the pole
$\mu _i \sim
A(x) (Y_i^+ - Y_j^-)$. This singular line is extended to the
semi-infinite line which is common for the  $i-th$, and
$j-th$ particle.
However, the considered in the following section simple example
shows that there exists also a second singular line
related to interaction of two particles.
It is out-going on the $Y_j^+$-sheet and  belongs to the
in-going $(-)$-sheet of the particle $i,$ $Y_i^-$ .

Therefore, each pair of the particles (ij) creates two opposite
oriented in the space (future directed) singular twistor lines,
pp-strings.
The field structure of this string is described by singular
pp-wave solutions (the Schild strings) \cite{BurTwi,BurPra}.

If we have k particles, then in general, for the each Kerr's
particle $2k$ twistor lines belonging to its PNC  will turn
into singular null strings.

As a result, one sees, that in addition to the well known
Kerr-Newman solution for an
isolated particle, there are series of the corresponding solutions
which take into account presence of the surrounding
particles, being singular along the twistor lines which are common
with them.

By analogue with QED, we call these solutions as `dressed' ones
to differ them from the original `naked' Kerr-Newman solution.
The `dressed' solutions have the same position and orientation as
the `naked' ones, and differ only by the appearance of
singular string along some of the twistor lines of the Kerr PNC.

\begin{figure}[ht]
\centerline{\epsfig{figure=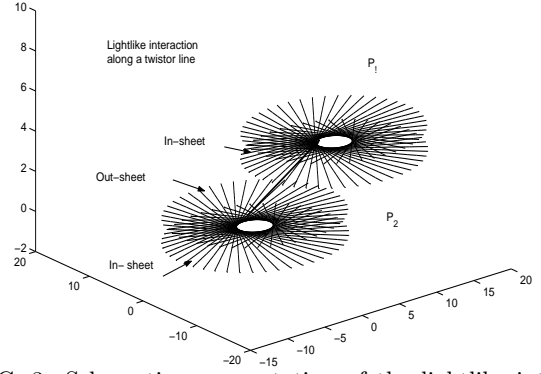,height=6cm,width=7cm}}
\caption{Schematic representation of the lightlike interaction
 via a common twistor line connecting  out-sheet of one particle
 to in-sheet of another.}
\end{figure}

\section{Example. Two-particle solution.}

Let us concretize solution for two particles $P1$
and $P2$.  Taking $F(Y)=F_1(Y) F_2 (Y)$, where

\bea F_1 =A_1(x)
(Y-Y_1^+)(Y-Y_1^-), \\
F_2 =A_2(x) (Y-Y_2^+)(Y-Y_2^-) , \label{F12}
\eea

one obtains that the `dressed' complex distance is

\bea \nonumber \tilde R =- d_Y F = - A_1 A_2 [ (Y-Y_1^-)(Y-Y_2^+)(Y-Y_2^-)
\\ \nonumber + (Y-Y_1^+)(Y-Y_2^+)(Y-Y_2^-)  \\
\nonumber +(Y-Y_1^-)(Y-Y_1^+)(Y-Y_2^-) \\
+(Y-Y_1^-)(Y-Y_1^+)(Y-Y_2^+)] . \label{dYF} \eea

If we are positioned on the sheet $2^+$, we have $Y=Y_2^+$, and
$\tilde R$ contains only one term,

\be \tilde R =\m_2 \tilde r_2^+ ,
\label{Rr2} \ee
where
\bea
\m_2 =A_1 (Y_2^+ -Y_1^-)(Y_2^+ -Y_1^+), \\
\tilde r_2^+ = - A_2 (Y_2^+ - Y_2^-) .\label{r2+}
\eea

One sees that $\m_2$ has the null at the set of space-time points where
$Y_2^+ =Y_1^-$ or $Y_2^+ =Y_1^+$. This set is a semi-infinite twistor line on
the sheet $Y_2^+$ consisting of the segment connecting particles P2 and P1, as
it is shown on FIG.3, and of the semi-infinite extension of this segment which
is the common twistor line of the sheets $Y_2^+$ and $Y_1^+$.

Let us consider two simple cases:

a/ the particles are in
the rest at the points $P1=(d,0,0)$ and $P2=(0,0,0)$, and the
spins are orientated along the z-axis,

b/the same, but the points $P1=(0,0,d)$, i.e. the both particles lie
on the z-axis.

Function $F_1(Y)$ and $
F_2(Y)$ may be used in the form (\ref{FiABC}) where coefficients
$A,B,C$ are determined by complex world lines $X^{(1)}_0(\t)$ and
$X^{(2)}_0(\t)$ of these particles \cite{BurNst,BurMag}.

In the null Cartesian coordinates  $(u, \ v, \ \Z , \ \z )$
they are given by the relations \cite{BurMag}

\bea A &=& (\Z  - \Z_0) \`v_0 - (v-v_0) \`\Z_0 , \\ \nonumber
B &=& (u-u_0) \`v_0 + (\z - \z_0
) \`\Z_0 - (\Z - \Z_0) \`\z_0 - (v - v_0) \`u_0 , \\ \nonumber
C &=& (\z -   \z_0 ) \`u_0 - (u -u_0) \`\z_0 . \eea

Particles $P1$ and $P2$ are described by complex world lines

{\bf Case a/ : }
\be    X^{(1)\m}_0 = \{\t,d,0,ia\}, \quad
 X^{(2)\m}_0 = \{\t,0,0,ia \}, \ee
or in the null coordinates

  \bea X^{(1)}_0 = 2^{-1/2}\{\t+ia, \ ia-\t, \ d, \ d \}, \\
 X^{(2)}_0 =  2^{-1/2}\{\t+ia, \ ia-\t, \ 0, \ 0 \}, \eea

{\bf Case b/ : }
\be    X^{(1)\m}_0 = \{\t,0,0,d+ia\}, \quad
 X^{(2)\m}_0 = \{\t,0,0,ia \}, \ee
and in the null coordinates

  \bea X^{(1)}_0 = 2^{-1/2}\{\t+d+ia, \ d+ ia-\t, \ 0, \ 0 \}, \\
 X^{(2)}_0 =  2^{-1/2}\{\t+ia, \ ia-\t, \ 0, \ 0 \}, \eea

Since particles are in the rest, we have for the both particles in the
both cases
$ \dot u_0=-\dot v_0 = 2^{-1/2} ,$ and
$ {\dot \z} _0={\dot \Z} _0 = 0 . $

Coefficients  for the functions $F_1$ and $F_2$ take the form:

{\bf Cases a/ and b/ :}

\bea A_2=- \frac 12 (x-iy), \\ \nonumber
B_2 = -z+ia, \\ \nonumber
C_2 = \frac 12 (x+iy).
\eea

{\bf Case a/ :}

\bea A_1=- \frac 12 (x-iy -d), \\ \nonumber
B_1 = -z+ia, \\ \nonumber
C_1 = \frac 12 (x+iy -d),
\eea

{\bf Case b/ :}

\bea A_1=- \frac 12 (x-iy), \\ \nonumber
B_1 = -z+d + ia, \\ \nonumber
C_1 = \frac 12 (x+iy),
\eea

The roots of the equations
$F_1=0$ and $F_2=0$ will be:

{\bf Case a/ :}

\be Y_1^\pm = \frac {ia -z \pm \tilde r_1}{x-iy-d}, \qquad
Y_2^\pm = \frac {ia -z \pm \tilde r_2}{x-iy} \label{Y12pma},
\ee
where
\bea
\nonumber
\tilde r _1=\sqrt{(x-d)^2 +y^2 +(z-ia)^2} , \\
\tilde r _2= \sqrt{x^2 +y^2 +(z-ia)^2} .
\label{tr12a}
\eea

One can find two singular twistor lines which are determined by
the equation $ Y_2^+ =Y_1^- $,  or
\be \frac {ia -z + \tilde r_2}{x-iy} =
\frac {ia -z - \tilde r_1}{x-iy-d}\label{commona}. \ee

It is enough to find at least one point for each of these lines. Looking
for such a point at $\tilde r_2 =0$, we obtain
\be r_1=d, \quad y=-a, \quad
x=z=0, \label{point1a}\ee which gives

\be Y_1^- = \frac {ia -d}{ia-d} =1,  \quad
Y_2^+ =\frac {ia }{ia} =1 ,  \ee
and the null direction corresponding $Y= Y_1^- =Y_2^+=1.$

This singular line is going via the point (\ref{point1a}) with null
direction $Y=1$. Another line is going via the point
\be \tilde r_2=0, r_1=-d, \quad y=a, \quad
x=z=0, \label{point2a}\ee
and has the opposite space direction $Y= Y_1^- =Y_2^+=-1$.
These two lines lie in the common equatorial plane of the both
particles and are tangent to the both singular rings.

{\bf Case b/ :}

Similar,

\be Y_1^\pm = \frac {ia +d -z \pm \tilde r_1}{x-iy}, \qquad
Y_2^\pm = \frac {ia -z \pm \tilde r_2}{x-iy} \label{Y12pmb},
\ee
where
\bea
\nonumber
\tilde r _1=\sqrt{x^2 +y^2 +(z-d-ia)^2} , \\
\tilde r _2= \sqrt{x^2 +y^2 +(z-ia)^2} .
\label{tr12b}
\eea
The equation   $Y_1^- =Y_2^+$ takes the form
\be
\frac {d+ia -z - \tilde r_1} {x-iy}=
\frac {ia -z + \tilde r_2} {x-iy}. \ee
Solutions are given by the equation
$\frac {d-\tilde r_1 - \tilde r_2}{x-iy}=0.$
Although, there appears indefinite limit by $x-iy \to 0,$
 one can show that
this expression tends to zero by $d-r_1 -r2=0$
and one obtains two singular lines which lie on the
axis $z$ and have the opposite directions $Y=0$ and
$Y=\infty$.\fn{Since particle P2 has the standard Kerr
position and orientation, one can use the known relations
for the Kerr angular coordinates,
$x-iy=(r_2-ia) e^{-i\phi} \sin \theta, \quad
z=r_2 \cos\theta$ and
$\tilde r_2= r_2 + ia \cos \theta ,$ which allows one to find the
limit by $x-iy\to 0.$}

Let us summarize the expressions for the `dressed' solution on the sheet
$i=2+$, where we will omit indices $2^+$ for coordinate $Y$

In the both cases it has the Kerr-Schild form
\be g^{(2)}_\mn =\eta _\mn + 2H^{(2)}P_2^{-2} e^{3}_{\mu} e^{3}_{\nu}
\label{g2} ,\ee
where

\be H^{(2)} = \frac {m_0}2 (\frac 1{\mu_2\tilde r_2} + \frac 1 {\mu_2^*\tilde
 r_2^*}) + \frac e{2 |\mu_2 \tilde r_2|^2} , \label{hKS2} \ee

Electromagnetic field  is determined by the
 two complex tetrad components
\be
\cF ^{(2)} _{12} = A^{(2)} (Z^{(2)} )^2
\ee
and
\be
\cF ^{(2)} _{31} = -(A^{(2)} Z^{(2)}),_1 \ ,
\ee
where for $\psi=e$ we have

\be  A^{(2)} =\frac e {(\mu _2 P_2)^2} .
\label{Aren2} \ee

Congruence is $e^3= P_2 k^{(2)}_\m dx^\m$ given by the form
\be
 e^3 = du+ \Y d \z  + Y d \Z - Y \Y d v ,
 \ee
where function $P_2$ is determined by the relation
$P_2=\dot X_0^{(2)\m} e^3_\m$ (see \cite{BurMag}) which gives for the
particle $P2$ in the rest the
known expression \cite{DKS}
\be P_2=2^{-1/2}(1+Y\Y).
\ee
Function $Y$ has the form
$Y=e^{i\phi} \tan \frac \theta 2,$ where $\phi$ and $\theta$ are
asymptotically the usual angular coordinates. However, their
relations to Cartesian coordinates is strongly deformed near singular ring,
so the dependencies $Y_i^\pm(x)$ turn out to be different for the different
sheets and rather complicate. They are nontrivial even for the Kerr's
particle $P2$ which has the standard position and orientation.
The corresponding function
\be Y\equiv Y_2^+ = \frac {ia -z + \tilde r_2}{x-iy}
\label{Y2p}, \ee where
\be \tilde r _2= \sqrt{x^2 +y^2 +(z-ia)^2 },
\label{tr2}
\ee
and one can get for particle $P2$ the known (Kerr-Schild) coordinate
relations
\be \tilde r _2  = r +ia \cos \theta ,  \ee
\be
x+iy =(r+ia)e^{i\phi}\sin \theta,  \quad
z=r\cos \theta.
\ee
Finally, the function $\m_2$ has the complicate form:

{\bf in the case a/ :}

\be
\m_2 = -\frac 12(x-iy -d) (Y -Y_1^-)(Y -Y_1^+), \label{mu2a}
\ee
where
\be Y_1^\pm = \frac {ia -z \pm \tilde r_1}{x-iy-d}, \ee
and

\be \tilde r _1= \sqrt{(x-d)^2 +y^2 +(z-ia)^2 },
\label{tr1a}
\ee

and {\bf in the case b/ :}

\be
\m_2 = -\frac 12(x-iy) (Y -Y_1^-)(Y -Y_1^+), \label{mu2b}
\ee
where
\be Y_1^\pm = \frac {ia + d -z \pm \tilde r_1}{x-iy}, \ee
and

\be \tilde r _1= \sqrt{x^2 +y^2 +(z-d-ia)^2 }.
\label{tr1b}
\ee

\section{Conclusion}
We considered the extended version of the Kerr theorem which,
being incorporated in the Kerr-Schild formalism,
allows one to get exact multiparticle Kerr-Schild solutions.

One of the principal new properties of these solutions is that
they are multi-sheeted. This property is a natural generalization
of the known two-sheetedness of the Kerr geometry and
 is related with multi-sheetedness of the corresponding
twistorial spaces of the geodesic and shear-free principal null
congruences.

Another new feature is the existence of the `naked'
and corresponding `dressed'
Kerr-Newman solutions, which resembles
the structure of dressed particles in QED\fn{Similar structure
appears also in DN sector of D-strings in superstring theory
\cite{Hul,BurOri}.}
The `naked' solution is the usual Kerr-Newman solution
for an isolated particle, while the `dressed' solutions are
the series of corresponding solutions, in which the selected
Kerr-Newman particle is surrounded by other particles.

Finally, the following from these solutions gravitational and
electromagnetic interaction of the particles via singular
pp-strings is also surprising and gives a hint that the photons
and gravitons may have the structure of pp-strings.

Moreover, the obtained multiparticle solutions are related to
multi-sheeted twistorial space-times, showing that a twistorial
web net of pp-string covers the space-time.

It suggests that the obtained multiparticle solutions may have
a relation to twistorial structure of vacuum which was conjectured
in many Penrose publications. It looks not too wonder,
since the multiplicative generating function of the Kerr
theorem (\ref{multi})  has been taken
in analogue with  the structure of higher spin gauge theory
\cite{Vas} and is reminiscent of a twistorial version of the Fock
space.

On the other hand, the Kerr-Newman solution has the double
gyromagnetic ratio $g=2$ as that of the Dirac electron,  which
raises the question on the relation of the Kerr spinning particle
to the Dirac electron \cite{DirKer}, as well as to the other
elementary particles. In these cases the relation $a>>m$ is
valid, and the black  hole horizons disappear providing
for the one-particle Kerr-Newman solution a
simple two-sheeted topology, where the analytic extension
of Kerr-Newman space-time may be represented by splitting
of the known conformal diagram, as it is shown in Fig.4.

\begin{figure}[ht]
\centerline{\epsfig{figure=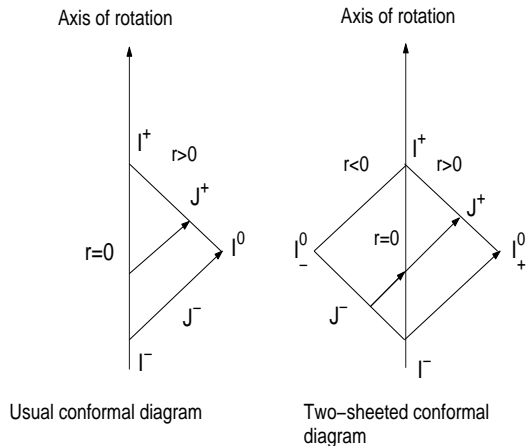,height=6cm,width=7cm}}
\caption{The usual conformal diagram of the flat Minkowski space-time (left
figure) and the twosheeted conformal diagram of the almost flat Kerr-Newman
 space-time with $a>>m$. The space-like infinity $I^0$ splits into
 $I^0_-,$ corresponding to $r=-\infty,$ and
 $I^0_+,$ corresponding to $r=+\infty,$.} \end{figure}

It allows one to conjecture that the presented multi-particle
solutions has a relation to the structure of spinning particle
and may
also shed some light on some problems of quantum gravity, which is also
supported by the old and modern expectations of the important
role of twistors in quantum theory \cite{WitNai,PenGra}.

The opposite case, $a<m$,  is important for
astrophysical applications. Although, the appearance of the
multisheeted structures in this case seems very
problematic,  the solutions with axial singular lines are actual,
since the semi-infinite singular strings may be related to
jet formation.  Treatment of the causal
structure of these solutions will be given in a separate paper
\cite{BEHM1}.  However, one can preliminarily note, that these
axial singularities change drastically the topological structure
of horizon, leading to formation of a hole in the event horizon.
So, the resulting `black hole' turns out to be not black.

\section*{Acknowledgments.} Author thanks G. Alekseev and the
participants of the seminar on Quantum Field Theory at
the Physical Lebedev Institute for very useful discussions.
This work was supported by the RFBR Grant
04-0217015-a.

\section*{Appendix A. Basic relations of the Kerr-Schild formalism}
Following the notations of the work
\cite{DKS},  the Kerr-Schild null tetrad
 $e^a =e^a_\m dx^\m $ is determined by relations:
\begin{eqnarray}
e^1 &=& d \zeta - Y dv, \qquad  e^2 = d \bar\zeta -  \bar Y dv, \nonumber \\
e^3 &=&du + \bar Y d \zeta + Y d \bar\zeta - Y \bar Y dv, \nonumber\\
e^4 &=&dv + h e^3,\label{ea}
\end{eqnarray}
and
\be g_{ab}= e_a^\m e_{b\m} = \left(
\begin{array}{cccc} 0&1&0&0 \\ 1&0&0&0 \\ 0&0&0&1 \\ 0&0&1&0 \end{array}
\right). \label{gab} \ee
Vectors
$e^3, e^4$ are real, and $ e^1, e^2 $ are
complex conjugate.

The inverse (dual) tetrad has the form
 \bea
  \d_1 &=& \d_\z  - \Y \d_u ;
\nonumber\\
\d_2 &=&  \d_\Z - Y \d_u ;
\nonumber\\
 \d_3 &=&  \d_u - h \d_4  ;
\nonumber\\
 \d_4 &=&  \d_v + Y \d_\z + \Y \d_\Z - Y  \Y \d_u ,  \label{1.10}
\eea
where $\d _a \equiv ,_a \equiv e_a^\m \d ,_\mu $.

 The Ricci rotation coefficients are
given by
\be \G ^a_{bc} = - \quad e^a_{\m;\n} e_b^\m e_c^\n.  \label{Gabc)}
\ee
 The PNC have the $e^3$  direction as tangent.
PNC is  geodesic if and
only if $\G_{424} = 0$ and shear free if and only if $\G_{422} = 0$.
 corresponding complex conjugate terms are $\G_{414} = 0$ and $\G_{411} = 0$.

The connection forms in this basis  are
\be \G_{42} = \G_{42a} e^a  = - d Y - h Y,_4
e^4 .  \label{1.11}
\ee

The congruence  $e^3 $ is geodesic if $ \G_{424} =
-Y,_4 (1-h) = 0, $ and is shear free if $ \G_{422} = -Y,_2 = 0.$

Therefore,  the conditions
\be
Y,_2 = Y,_4 = 0,  \label{1.12}
\ee
define a shear-free and geodesic congruence.

\section*{Appendix B. Proof of the Kerr Theorem for the
Kerr-Schild background.}

Starting from the geodesic and shear free conditions
$Y,_2 =  Y,_4 = 0 $,
one obtains the differential of function $Y$  in the form
\be
d Y = Y,_a e^a  = Y,_1 e^1 + Y,_3 e^3.
  \label{2.1}\ee

$Y,_1 =Z$ is an important parameter: complex expansion of congruence,
$Z=\rho +i \omega, $ where  $\rho=expansion$ and $\omega = rotation$.

One needs to work out the form  $Y,_3$. By using relations
(\ref{1.10}) and their commutators one finds
 \be   Z,_2 = (Z - \bar Z) Y,_3.
  \label{2.2}\ee
Straightforward differentiation of $Y,_3$ gives the equation
\be
Y,_{32} = ( Y,_3) ^2    .
\label{2.3}\ee
Then, by using (\ref{2.2}) and (\ref{2.3}) one obtains the equation
\be
(Z^{-1} Y,_3),_2 = \bar Z ( Z^{-1} Y,_3 )^2.
\label{2.4}\ee

This is a first-order differential equation for the function
$Z^{-1} Y,_3$.
Its general solution can be obtained by substitution $x=Z(Y,_3)^{-1}$
and has the form
\be
Y,_3 = Z ( \phi - \bar Y) ^{-1},
\label{2.5}\ee
where $\phi$ is an arbitrary solution of the equation $ \phi,_2 =0$.
Analogously, by using the relation $Y,_{34}=-Z Y,_3$ one obtains
$\phi,_4=0$, and therefore $\phi$ may be an arbitrary function satisfying
\be
\phi,_2=\phi,_4 =0.
\label{2.6}\ee
One can easily check that the three projective twistor coordinates
$ \l_1 = \z - Y v, \qquad \l_2 =u + Y \Z,$ and $ Y$
satisfy the similar relations $ (.),_2=(.),_4 =0$.

Since the surface
$\phi=const$ forms a sub-manifolds of  $CM^4$ which has the complex dimension
three, an arbitrary function $\phi$ satisfying (2.6) may be presented as
function of three projective twistor coordinates $\phi=\phi(Y, \l_1, \l_2)$.
Now we can substitute $Y,_3$ in (\ref{2.1}) that implies
\be
Z^{-1} (\Y - \phi) d Y = \phi
(d \z -Y d v) + (du + Y d \Z).
\label{2.7}\ee
If an arbitrary holomorphic function $F (Y,\l_1,\l_2)$ is given, then
differentiating the equation $F (Y,\l_1,\l_2)=0$
 and comparing the result with (\ref{2.7}) one finds that
\be PZ^{-1}= - \quad
 d F / d Y  , \label{FY} \ee
where
\be
P = \d_{\l_1} F - \Y \d_{\l_2} F,
\label{2.8}\ee
where the function $P$ can also be defined as
\be
P = (\phi - \b Y)\d_{\l_2} F.
\label{2.9}\ee

Therefore,  we have:

{\bf Corollary 1:}

For arbitrary holomorphic function  of the projective
twistor variables $F (Y,\l_1,\l_2)$, the equation $F=0$ determines
function $Y(x)$ which gives the congruence of null directions $e^3,$
(\ref{ea}) satisfying the geodesic and shearfree conditions $Y,_2=Y,4=0.$

{\bf Corollary 2:}

Using (\ref{1.10}), one sees that the explicit form of the geodesic and
shearfree conditions $Y,_2=Y,4=0$ is
$(\d_\Z - Y \d_u)Y =0$ and  $ (\d_v + Y \d_\z) \d_u)Y =0$. It does not
depend on function $h$ and coincides with these conditions in Minkowski
space. Therefore, the resulting PNC is geodesic and shearfree with respect
to the Kerr-Schild background as well as with respect to the auxiliary
Minkowski metric $\eta _\mn$.

{\bf Corollary 3:}

Function $F$ (which we call generating function) determines two
important functions
\be
P = \d_{\l_1} F - \Y \d_{\l_2} F,
\label{P}\ee
and
\be PZ^{-1}= - \quad
 d F / d Y  , \label{FYC} \ee

{\bf Corollary 4:}

Singular points of the congruence, where the complex divergence
$Z$ blows up, is defined by the system of equations

\be F=0,\quad dF/dY =0  \label{sing}\ee

Note, that
\be
\tilde r =PZ^{-1}. \label{trZ}
\ee
is complex radial distance which is related to complex
representation of the Kerr geometry \cite{BurNst}.

{\bf Corollary 5:}

The following useful relations are valid
\be
\bar Z Z^{-1} Y,_3=- (\log P),_2 \ , \quad P,_4=0 .
\label{2.11}\ee

{\bf Proof.}
So far as $\d_2 \d_{\l_2}F =0$, one sees that
\be
(\log P),_2 = -\bar Z (\phi -\Y)^{-1},
  \label{2.10}\ee
then (\ref{2.5}) leads to first equality of (\ref{2.11}).
The relation $P,_4=0$ follows from (\ref{P}) and properties of
the twistor components $Y,_4= (\l_1),_4 =  (\l_2),_4 =0$.

\section*{Appendix C. Integration of the Einstein-Maxwell field
equations for arbitrary geodesic and shearfree PNC}

Following \cite{DKS}, the Einstein-Maxwell field equations written
in the above null tetrad form become:

\be R_{ab} = -2(F_{ac}{F_b}^c -\frac 14 F_{cd}F^{cd}), \label{grav}
\ee

\be {{\cal F}^{ab}}_{;b}= {{\cal F}^{ab}}_{,b} +
{\G ^a}_{cb}{\cal F}^{cb} +
{\G ^b}_{cb}{\cal F}^{ac} =0 ,\label{Max}
\ee
where electromagnetic field is represented by complex tensor
\be
{\cal F}_{ab} =- {\cal F}_{ba} = F_{ab} +\frac i2 \eta _{abcd} F^{cd},
\label{cF}
\ee
and $\eta _{abcd}$ is completely skew-symmetric and
\be
\eta _{1234} =i .
\ee

The geodesic and shearfree conditions $Y,_2=Y,_4=0$ reduce strongly
the list of gravitational and Maxwell equations. As a result,
one obtains for the tetrad components
\be R_{24} =R_{22} =R_{44}=R_{14} =R_{11} = R_{41} =R_{42} =0.
\label{(5.1)}\ee
The equation $ R_{44} =0$  gives $F_{42} =F_{41}=0 ,$ which simplifies
strongly ${\cal F}_{ab},$ leaving only two nonzero complex components

\be {\cal F}_{12} = {\cal F}_{34} =  F_{12} + F_{34}
\label{cF12}, \ee
and
\be {\cal F}_{31} = 2 F_{31}
\label{cF31}. \ee

The survived gravitational equations are
\be (Z+\bar Z) [ h,_4 + (\bar Z -Z)h] +2 Z^2 h=
{\cal F}_{12}{\bar {\cal F}}_{12}, \label{g14} \ee
and
\be [ h,_4 + (\bar Z -Z)h],_4 + 2 Z [ h,_4 + (\bar Z -Z)h]=
-{\cal F}_{12}{\bar {\cal F}}_{12}. \label{g24} \ee

Adding these equations, one obtains the equation for $h$

\be h,_{44} + 2 (Z+\bar Z) h,_4 + 2Z\bar Z h=0  \label{h44} \ee
which has the general solution
\be h= \frac 12 M (Z+\bar Z) + B Z\bar Z, \quad M,_4=B,_4=0.
\label{hMB}\ee
By obtaining this solution, the very useful relation
\be Z,_4 = -Z^2 \label{Z4}\ee was used.\fn{It was obtained
from commutation relation for $Y,_{[14]}$.}
In the same time, one of the Maxwell equations
\be {\cal F}_{12},_4 + 2 Z F_{12} =0
\label{dF12} \ee  may easily be integrated, leading to
\be {\cal F}_{12} =AZ^2, \quad A,_4=0 .\ee
This equation, being inserted in (\ref{g24}) yields
$B=-\frac 12 A\bar A,$
which gives the final general form for $h$ for any geodesic and
shearfree PNC
\be h= \frac 12 M (Z+\bar Z) + A\bar A Z\bar Z.
\label{hdks}\ee
Other Maxwell equations take the form

\be {\cal F}_{31} = \gamma Z -(AZ),_1 , \quad \gamma ,_4=0  \ ,
\label{F31} \ee
\be A,_2 - 2Z^{-1}\bar Z Y,_3 A=0, \label{A2} \ee

\be A,_3 - Z^{-1} Y,_3 A,_1 -\bar Z^{-1} \bar Y,_3 A,_2
+\bar Z^{-1} \gamma,_2 - Z^{-1} Y,_3 \gamma =0. \label{A3} \ee

And finally, after teddious calculations, two last gravitational equations
may be obtained
\be M,_2 - 3 Z^{-1} \bar Z Y,_3 M -A\bar \gamma \bar Z =0,
\label{M2} \ee
and
\be M,_3 - Z^{-1} Y,_3 M,_1 - \bar Z^{-1} \bar Y,_3 M,_2
-\frac 12 \gamma\bar\gamma =0.
\label{M3} \ee

These equations may further be reduced for $\gamma=0 ,$
which means the restriction by a stationary electromagnetic
field, without wave excitations. It yields

\be {\cal F}_{31} = -(AZ),_1 ,
\label{F310} \ee
\be A,_2 - 2Z^{-1}\bar Z Y,_3 A=0, \label{A20} \ee

\be A,_3 - Z^{-1} Y,_3 A,_1 -\bar Z^{-1} \bar Y,_3 A,_2 =0.
\label{A30} \ee

\be M,_2 - 3 Z^{-1} \bar Z Y,_3 M  =0,
\label{M20} \ee
and
\be M,_3 - Z^{-1} Y,_3 M,_1 - \bar Z^{-1} \bar Y,_3 M,_2 =0,
\label{M30} \ee

\be M,_4=A,_4=0 \label{MA0} .\ee

These equations contain the function $Z^{-1} Y,_3$ which is determined
by the Consequence 5 of the Kerr theorem, (\ref{2.11}),
as follows
\be
 \bar Z Z^{-1} Y,_3=- (\log P),_2 \ , \quad P,_4=0 .
\label{P2}\ee

So, by using this expression one obtains

\be (\log AP^2),_2  =0, \quad A,_4=P,_4=0
\label{AP} \ee

\be (\log M P^3),_2 =0, \quad M,_4=P,_4=0.
\label{MP}
\ee
Since $Y,_2=Y,_4=0,$ the general solutions of these equation
have the form
\be A= \psi(Y)/P^2
\label{Apsi} \ee
and
\be M= m(Y)/P^3.
\label{Mm} \ee

\section*{Appendix D. Multisheeted PNC and double twistor
bundles} In general case, the twistorial structure for an i-th particle
is determined by solution $Y_i^\pm (x)$ and forms a double
twistor bundle $E =CM^4\times CP^1,$ where
 $CP^1=S^2$ is the Riemann sphere, parametrized by projective
angular coordinate $Y\in CP^1,$ and $CM^4$ is a complexified
Minkowski space-time:

\bea
           \  \  \ \  &E^5 & \ \ \ \ \  \\ \nonumber
           \pi _2 \ \swarrow  & \ \ &   \searrow \ \pi _1  \\ \nonumber
           CP^3 \ \ \ \ & \ \ & \ \ \ \ CM^4  .   \label{twbundle}
\eea
The bundle with the base
$CM^4$ has the fiber $F_x =\pi _1 ^{-1}(x)= CP^1=S^2, $ in which
the complex coordinate $Y\in CP^1$ parametrizes
the subspace of complex null $\alpha$-planes incident to
the point $x.$

The  base of dual bundle, $CP^3$, is the space of projective
twistor coordinates
$Z=
\{ Y,\quad \l_1=\z - Y v, \quad \l_2=u + Y \Z \} \in CP^3 \}.$
The fiber of this bundle $F_{Z} =\pi _2 ^{-1}(Z)$ represents a
fixed complex null plane (twistor) in $CM^4$.

The subset $PN$ of ``null'' twistors
 $Z_N=\{ Y,\quad \z - Y v, \quad u + Y \Z \}\in PN \subset CP^3,$
 is determined by the coordinates
 $\{ Y,\quad \l_1=\z - Y v, \quad \l_2=u + Y \Z \},$ where
$(u,v,\z,\Z)$ are the null Cartesian coordinates of the real
Minkowski space-time $M^4$, i.e. $u$ and $v$ are real, and
$\z,$ and $\Z$ are complex conjugate.
The subset $PN$ may also be selected by quadric
\be PN = \{Z : Z_\alpha \bar
Z^\alpha =0 \}\label{PN}, \ee
$PN\in CP^3 \times \bar {CP}^3.$

Therefore, the above double twistor bundle may be restricted to
the real twistor bundle:
$\quad E = M^4 \times CP^1$
\bea
           \  \  \ \  &E^5 & \ \ \ \ \  \\ \nonumber
           \pi _2 \ \swarrow \ & \ \ &  \ \searrow \ \pi _1  \\ \nonumber
           PN \ \ \ \ & \ \ & \ \ \ \ M^4  ,   \label{rebundle}
\eea
where the base $M^4$ is the real Minkowski space, and the fiber
is again the Riemann sphere $CP^1$, parametrized by complex
projective angular coordinate $Y$.
The fiber of this bundle $F_x =\pi _1 ^{-1}(x)=CP^1$ is the
subspace of the real null rays incident to the point $x\in M^4.$

The dual bundle has the base  $PN$
which is  parametrized by the null projective twistors which lie
on the quadric (\ref{PN}),  $Z \in PN ,$ and
the  fiber of this bundle $F_{Z} =\pi _2 ^{-1}(Z)$ represents a
fixed null ray (real twistor) in $M^4$.
 One can also note that $Y_i
(x)=Y_j(x)$ on the common (ij)-twistor line. Since the tetrad
vectors are determined by the values of $Y, \ \bar Y$, it
 has a
consequence that all the tetrad vectors $e^a$ match on this
line for the different sheets over $M^4$, and also over
${\cal K}^4$ for exception the singular lines.
It confirms that the sheets are connected, forming the covering
space over $CP^1$ with the set of branch points
which are determined by the equations (\ref{singlines}).
Since these null rays are determined by the tetrad null vector
 $e^3(Y,\bar Y)$ at the point $x$, this twistor bundle may be
 extended
 to the cotangent bundle of the Kerr-Schild null tetrad $e^a, \
a=1,2,3,4$ which is also fixed by the fields of $Y(x)\in CP^1$ on
the base space $M^4$ (See Appendix A).
 On the other hand,
the twistor bundle $E=CP^1 \times M^4$ may be
extended to the complexified Minkowski space, $E=CP^1 \times
CM^4$, with base $CM^4\ni x$ and fibers $F_x$ which are incident
to $x$ complex null planes ($\alpha$-planes), spanned by the
tetrad vectors $e^1 \wedge e^3$ and controlled by $Y\in CP^1=S^2.$

PNC, or twistorial structure, related to i-th particle is a
section of fibre bundle $s_i(E)=s(E,Y_i(x))=s\{x,e^3(Y_i(x))\}$.
Multisheeted twistor space is formed by the set of
sections $s_i(E)=s(E,Y_i(x))$,
which are controlled by multisheeted function $Y(x)$.

\end{document}